\documentclass[journal]{IEEEtran}

\usepackage{url} 
\usepackage{nameref} 
\usepackage{graphicx} 
\usepackage{multirow} 
\usepackage{cite}
\usepackage{amsmath} 
\usepackage{flushend} 

\newcommand*{\secref}[1]{Section~\ref{#1}}
\newcommand*{\figref}[1]{\figurename~\ref{#1}}
\newcommand*{\tabref}[1]{Table~\ref{#1}}

\newcommand{\figwidth}{3.5in}

\begin{document}

\title{Rapid Skill Capture in a First-Person Shooter}

\author{David~Buckley,
        Ke~Chen,
        and~Joshua~Knowles%
\thanks{This work has been submitted to the IEEE for possible publication. Copyright may be transferred without notice, after which this version may no longer be accessible.

This work was supported by the Engineering and Physical Research Council [EP/I028099/1].

D. Buckley, K. Chen and J. Knowles are with the School of Computer Science, University of Manchester, Manchester M13 9PL, U.K. (e-mail: david.buckley@cs.man.ac.uk; ke.chen@manchester.ac.uk; j.knowles@manchester.ac.uk).}}

\markboth{Manuscript for the IEEE Transactions on Computational Intelligence and AI in Games}%
{Buckley \MakeLowercase{\textit{et al.}}: Short-Term Skill Capture in a First-Person Shooter}


\maketitle

\begin{abstract}


Various aspects of computer game design, including adaptive elements of game levels, characteristics of `bot' behavior, and player matching in multiplayer games, would ideally be sensitive to a player's skill level.
Yet, while difficulty and player learning have been explored in the context of games, there has been little work analyzing skill per se, and how it pertains to a player's input.
To this end, we present a data set of 476 game logs from over 40 players of a first-person shooter game (\emph{Red Eclipse}) as a basis of a case study. We then analyze different metrics of skill and show that some of these can be predicted using only a few seconds of keyboard and mouse input.
We argue that the techniques used here are useful for adapting games to match players' skill levels rapidly, perhaps more rapidly than solutions based on performance averaging such as TrueSkill.

\end{abstract}

\begin{IEEEkeywords}
First-person shooter, player modeling, skill capture, skill metrics, skill prediction.
\end{IEEEkeywords}

\section{Introduction}
\label{sec:intro}

\IEEEPARstart{S}{kill} is an important component of any recreational or competitive activity. Not only does it contribute to the result, the relationship between skill and difficulty of the activity affects the experience of those taking part. Players in a game, for instance, often have as little fun beating novices as they do being dominated by highly accomplished players.

In our research, skill is a property of a player, defined in terms of their average performance. This discounts notions of `skillful' behavior other than those that aid in winning the game. The definition used here falls in line with existing skill metrics \cite{elo,trueskill}, and allows skill to be explicitly measured.

If a player's skill were known before they played, their opponents could be selected in a way that would optimize their experience of the game. In competitive games, this is known as matchmaking, and is widely used in online gaming. Single player games, on the other hand, use \emph{Dynamic Difficulty Adjustment} (DDA) \cite{dda:erl,dda:ai}, where the game's difficulty is changed according to the player's progress. \emph{Left 4 Dead}'s AI director is an example of this in action \cite{dda:l4d:director}.

Unfortunately, there is currently no quick and accurate way of measuring a player's skill. Bayesian methods, such as TrueSkill \cite{trueskill}, require several games before converging, depending on the number of players, and DDA relies on heuristic methods which are not necessarily representative of a player's skill \cite{dda:erl}. In a domain where a single bad experience can be enough to alienate someone, two or three games can be too many, so we seek to reduce this to a single game or less.

Whereas a player's performance may depend on several factors, including their opponents, their input, e.g. mouse and key presses, is consistent over several games. It is intuitive to assume that a skilled player will interact with the controls differently to a novice \cite{style:lda}. Instead of relying on performance as a metric for each player, we therefore consider using their input.

Towards this goal, we have performed a systematic study based on \emph{Red Eclipse}, a first-person shooter game (FPS). Game logs were automatically recorded during the study, storing input events, some game events and a few common measures of performance. In order to understand these measures, we present a thorough analysis of them and the features extracted from the input events. Building on the success of random forests in previous work \cite{skill:fps:input}, we then predict the player's skill with reasonable accuracy from only 10 seconds of data (see \figref{fig:speed:regression}).

Our main contributions can be summarized as follows: 1) a complete data set of games containing player input and results, 2) an investigation of the data set, validating a number of skill metrics and exploring their connection to input, and 3) a model capable of predicting a player's skill from less than a single game.

The rest of this paper is organized as follows. After a review of previous work in \secref{sec:background}, the data set is described in depth in \secref{sec:data}. We use the techniques presented in \secref{sec:method} to analyze different skill measures and player behavior in Sections \ref{sec:measures} and \ref{sec:features}. We finally present the skill prediction in \secref{sec:prediction}, discuss the implications of this research in \secref{sec:discussion}, and present our concluding remarks in \secref{sec:conclusions}.

\section{Previous Work}
\label{sec:background}

We define \textbf{skill} as the \emph{average level of performance over a set of games}. A value of skill only holds meaning for a particular set and for a particular averaging technique. This definition does not consider concept drift or learning, and assumes skill is averaged over a reasonable length of time.

The definition of skill used here is distinct from the term \textbf{ability} defined by Parker and Fleishman \cite{ability}: ``Ability refers to a more general, stable trait of the individual which may facilitate performance in a variety of tasks. \ldots The term skill is more specific: it is task oriented.''

\begin{figure}[!t]
\centering
\includegraphics[width=2.5in]{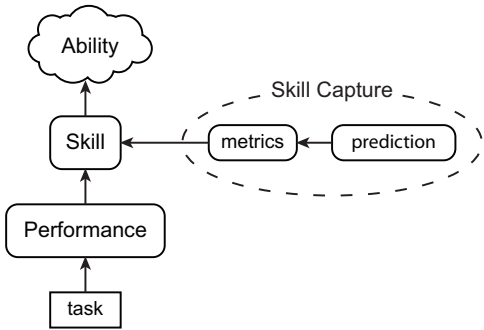}
\caption{The connections between the concepts used in this paper.}
\label{fig:terms}
\end{figure}

\textbf{Performance} is \emph{the value assigned to a person after a particular task has been completed}. This value, or measure, is defined by a \textbf{metric}, where different metrics may yield different performance measures for the same task and the choice of metric used affects the rankings of players within a game. The connection between skill and performance has been illustrated in \figref{fig:terms}, and is similar to the connection Chomsky draws between competence and performance \cite{chomsky:competence}.

We differentiate between a \textbf{skill metric}, which is calculated by averaging performance over time, and \textbf{skill prediction}, the process of predicting a skill metric using less information than that required by the metric. Thus, while the prediction may share the same unit as the metric, it is not guaranteed to produce the same ranking. Although both are considered methods of \textbf{skill capture}, this research assumes that a skill metric always has higher validity than a prediction.

\subsection{Performance and Skill Metrics}

There are numerous ways to measure performance of a task, and each video game has its own common metrics that are used by developers or its community. \emph{StarCraft} and \emph{Counter-Strike}, for instance, use win-loss metrics to determine the winner, whereas players of each game use \emph{actions-per-minute} and \emph{kill-to-death ratio} respectively to compare themselves. These can, and often are, averaged to provide players with skill metrics.

A common problem with metrics is `inflation', where players change their gameplay to manipulate their performance (and consequently their skill measure), contrary to how the developers intended them to play. Combining and adjusting different metrics is done in order to encourage desired behavior \cite{skill:ranking:bayes}. The WN6 algorithm used for \emph{World of Tanks}, for example, takes a variety of metrics and combines them using weightings and a series of mathematical operations to produce a single skill metric \cite{wn6}.

TrueSkill, unlike the simple metrics previously mentioned, averages performance using Bayesian updating \cite{trueskill}. The model, which is based on the Elo rating \cite{elo}, actually represents a belief in a player's skill, which can be reduced to produce a skill metric. The model uses rank as its performance metric, and can therefore cope with multiple teams of varying player sizes. The main criticisms of TrueSkill are its time to convergence, which can take several games to find a confident representation, and that values cannot be compared across different leagues \cite{trueskill:through-time}.

\subsection{Skill Prediction}

Skill metrics have the distinct disadvantage that performance measures must be taken over a set period of time in order to determine an average. Users of the TrueSkill algorithm, for instance, need to play anywhere between 3 and 100 games, depending on the number of players in each game. Skill prediction techniques seek to determine an individual's skill in significantly less time.

Kenneth Regan \textit{et al.} extend a chess end-game performance metric \cite{skill:chess:endgame} to complete chess games \cite{skill:chess:prediction}. Using the assumption that computers can play better than humans, a player's move is compared with those of a computer to produce a prediction of the player's performance. The authors then use Bayesian averaging over several moves in order to produce a skill prediction.

The task of skill prediction is not limited to games, and also extends to domains such as teleoperations \cite{skill:robots:hmm} and Human Computer Interaction (HCI) \cite{skill:hci:taskmodel,skill:hci:lowlevel}. The work in HCI takes advantage of a user's mouse input to predict their skill for a specific task and the system as a whole. Several useful features of the mouse are highlighted in this work, and were used for our own research. However, this work in HCI focuses on a predefined task with specific instructions that the users can learn very quickly. This contrasts with the task used in our own experiments, which is more analogous to `system skill'.

Within the domain of video games, there have been a few attempts at skill prediction, using a variety of techniques, including physiological monitoring, recording game events, and logging player input from the hardware. The first of these, monitoring physiological responses, explored skill in a fighting game \cite{skill:fighting:physiological}. The researchers distinguished between players of different skill using the performance metric `success rate' when inputting commands. However, while the work provides a foundation for further research, there was a very small number of participants and little analysis of the differences between player types. Moreover, physiological data collection can be intrusive, potentially distancing players from immersion, thus changing how they play.

An alternative to physiological data is using information about the game and high-level game events. This sort of data is easy to collect, and useful for other methods of prediction \cite{adapting-mario}. Mahlmann \textit{et al.} consider this data for predicting completion time in \emph{Tomb Raider: Underworld} \cite{predicting-tombraider}, a reasonable metric of performance for single-player games. The main focus of the paper was not on player skill, however, and the results of prediction were inconclusive.

Finally, the most closely related research was done in the real-time strategy (RTS) game \emph{StarCraft II} \cite{skill:rts:input}. In this work, Tetske \textit{et al.} successfully predict a player's skill level using `actions', the interactions between the player and the interface, from a substantial data set. Rather than predicting a skill metric, however, the model is trained to predict the league or group of each player, making the assumption that these categories accurately indicate skill. The research also makes no use of hardware input events, which along with skill metrics, are explored in depth in our own research.

\section{Data Set}
\label{sec:data}

To our knowledge, there does not exist a publicly available data set that specifically concerns `player input': the players' input to a game through the means of hardware, e.g. a mouse or keyboard. This paper therefore presents a substantial data set of game logs recorded from many different players of an FPS.

Designed for balance and representation of different player types, the data, and how it was collected is described here. The data set, scripts for manipulating it and further information can be found on our website\footnote{\url{http://www.cs.man.ac.uk/~buckled8/shortterm.html}}.

\subsection{Red Eclipse}

\begin{figure}[!t]
\centering
\includegraphics[width=\figwidth]{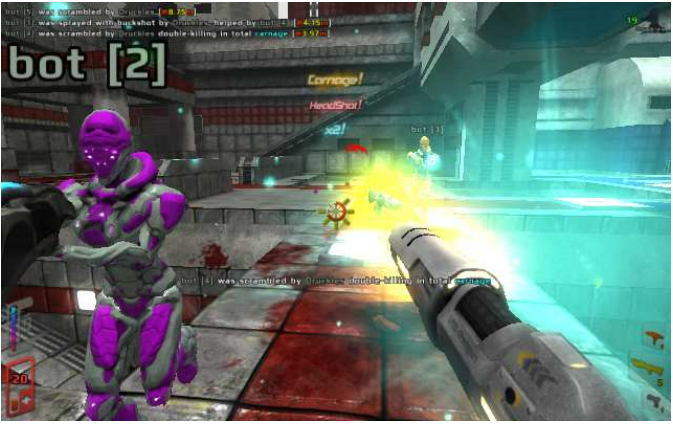}
\caption{A screenshot of the game used in our study, \emph{Red Eclipse}.}
\label{fig:red-eclipse}
\end{figure}

The test-bed for this experiment was an open-source first-person shooter, \emph{Red Eclipse}\footnote{\url{http://www.redeclipse.net}}, which is a fully-customizable, fast-paced action game that includes many common game mechanics from the FPS genre. A screenshot of the game can be seen in \figref{fig:red-eclipse}.

While \emph{Red Eclipse} strives to emulate traditional game mechanics, it also provides a `parkour' system, which is not present in most first-person shooters. The system allows players greater freedom in moving around their environment, but adds a further level of complexity. Many players tried to use this feature, but very few used it consistently.

The data collected from the games were limited to logging the inputs of the player and some information about the game. A timestamped log file was constructed for each game, recording the game's settings and a selection of events, including keyboard and mouse events and some game features such as kill and damage events.

\emph{Red Eclipse} allows users to modify the game settings in order to customize their experience. This includes the type of game they play (the game mode), the arena in which they play (the map), and the difficulty of simulated enemies (bots).

The game mode was set to \emph{deathmatch}, in which players compete to kill each other for the most points. This limited the complexity of rules and tactics used, and meant players were not dependent on the skill of their teammates. Each game was also set to three minutes; considered long enough for players to become immersed, but short enough to meet our goal of short-term skill capture.

Eight different maps were chosen in order to represent a range of playing environments. Some maps were more difficult for players, whereas others were harder for the bots. Six ranges of bot difficulty were used (40--50 to 90--100) defining the minimum and maximum difficulty. From a given range, inclusive of the two limits, the engine randomly selects an integer for each bot which defines its skill for that game.

\subsection{The Log File}

Although the structure of the log file was designed independently, inspiration was drawn from similar research being done at the time \cite{wild}. Each log file has a set of metadata that describes the game and a variable-length list of events. The log files, originally text-based, have been published as \texttt{JSON} objects. This is for flexibility and human-readability.

Each game comes with information that describes its settings. The list of metadata can be found in \tabref{tab:metadata} along with a brief description. The Client Number is set by \emph{Red Eclipse} when connecting to an online game, but is always 0 in this data set. Although the bot difficulties had a larger range, they were restricted to 40 and 100 in this experiment, as difficulties lower than 40 were considered minimally different.

\begin{table*}[!t]
\renewcommand{\arraystretch}{1.3}
\caption{The meta data for each game.}
\label{tab:metadata}
\centering
\begin{tabular}{ l | p{3.0in} | l }
\hline
Name      & Description & Example \\
\hline
Game ID   & A unique identifier for the game. & 127 \\
Player ID & A unique identifier for the current player. & 26 \\
Client Number & The number assigned to the player by the game. & 0 \\
Game Number & From the set of games played by one player, the position this game appears (starting from 0). & 5 \\
Map Name  & The name of the map that was selected for this game. & wet \\
Bot Min   & \multirow{2}{3.0in}{Each bot's difficulty is chosen randomly from between Bot Min and Bot Max. Possible values range from 0 to 101.} & 60 \\
Bot Max   & & 70 \\
Connect time & The time the user connected to the game (ms). & 1 \\
Disconnect time & The time the game ended (ms). & 185010 \\
Scoreboard & The final scoreboard for the game, including number of points and kills for each player (given by their client number). & {0: {'points': 8, 'kills': 3} ...} \\
Date \& time & The date the game was played and the time it started. & 2013-02-26, 14:40:54 \\
\hline
\end{tabular}
\end{table*}

Two types of events were extracted from the game: input events and game events. Input events were further separated into \emph{key presses}, \emph{mouse button presses} and \emph{mouse motion}. Keyboard and mouse button events contain a key identifier, the final state of the button and the action the button caused in the game. Mouse motion events have an x and y value (the number of pixels the mouse was moved), and were triggered roughly once every three milliseconds while the mouse was in motion.

The second category of events is a simplified summary of game events. These events, generated by the game, only concern events that happen to the player; in other words, interactions between bots is not considered. The events were chosen with the consideration of skill as a focus of the experiment.

\subsection{Data Collection}

The data set was compiled from an in-house experiment. This level of control gave both consistency and reliability to the data set. It also allowed the experimenters to ensure the data set remained balanced throughout.

Although the terms participant and player can be used interchangeably, we have attempted to attribute participant to the context of the experiment, and player to the context of the game.

\begin{figure}[!t]
\centering
\includegraphics[width=\figwidth]{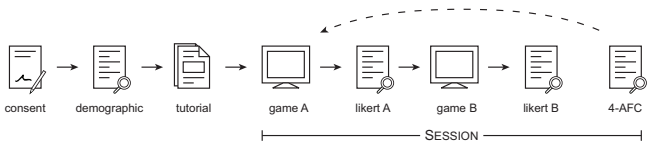}
\caption{The overall format of the experiment.}
\label{fig:experiment}
\end{figure}

The overall format for the experiment is presented in \figref{fig:experiment}. Each participant started by completed a demographic questionnaire at the start. They were then presented with a written tutorial and given as much time as they needed to read through it. This included a summary of general first-person shooter mechanics and more specific details about \emph{Red Eclipse}. Participants were allowed to ask questions at any point through the experiment or refer back to the tutorial, but the experimenter did not provide information voluntarily.

The main part of the experiment was split into `sessions', where a single session consists of a pair of games and a respective set of questionnaires,  as in \figref{fig:experiment}. A participant was allowed to complete as many sessions as they wanted. After each game, the participant answered questions about their experience, and at the end of each session, the participant would compare the two experiences. The questionnaires are described in the next section.

All participants used the same keyboard and mouse, and a headset was provided to wear at their discretion. The researcher was present in the room throughout the experiment to guide participants through the process and answer any questions. On three occasions, the researcher had to intervene to ensure participants followed procedure. For each of these games, there is roughly an 18s gap of missing game data. These games are highlighted on the website.

Finally, it is worth noting that the data, while only spanning a few weeks, is separated by several months. After the initial study \cite{skill:fps:input}, a further period of data collection was held in order to improve on existing problems with the data set. In particular, the second period was designed to correct imbalances of content, increase the overall number of games, and increase the number of games per player. From all 45 participants, 14 took part exclusively in the first period, 11 in the second and 20 took part in both periods.

\subsection{Questionnaires}

There were three different questionnaires used in total throughout the experiment: a demographic questionnaire, an experience-based questionnaire using the Likert scale \cite{likert}, and an experience-based questionnaire using 4 Alternative Forced Choice (4-AFC) \cite{afc}.

The demographic questionnaire was presented to participants before they started. This questionnaire gleaned information such as age, gender and, most notably, two self-reported measures of skill. The first measure, how many hours the participant plays per week, is a common question in research \cite{motivation:minecraft,corpus:mario}. The second, the number of first-person shooters played, was conceived in order to discount the effect of other genres, and account for the player's entire gaming experience, rather than playing habits. These questions were designed to be objective and avoid self-assessment, which players are notoriously poor at \cite{self-assessment}.

The two experience-based questionnaires used the same questions in two different forms. The first was Likert, to allow the participant to rate each game separately, and the second 4-AFC, comparing the last two games. There are advantages and disadvantages to each method, which are discussed more thoroughly in \cite{likert-afc}. Each of these questionnaires had four questions concerning the fun, the frustration, the challenge and the player's impression of the map. The first three questions have been used previously with some degree of success \cite{adapting-mario,dda:two-player}. In our research, the Likert questionnaire was worded as follows:

\begin{itemize}
  \item How much would you want to keep playing the game?
  \item How frustrating did you find the game?
  \item How challenging did you find the game?
  \item How lost did you feel while playing the map?
\end{itemize}

The first question, regarding fun, was chosen to allow players to question their current state of feeling, rather than remembering how they felt during the game. This was to mitigate the effects of memory on self-reported affect \cite{moment-based}.

\subsection{Data Distribution}

\begin{figure}[!t]
\centering
\includegraphics[width=\figwidth]{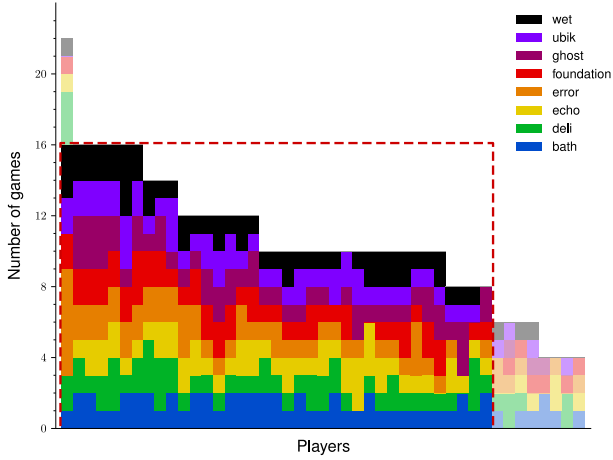}
\caption{The number of games played by each player. Games highlighted by the dashed box are those used in this research. Colors indicate which maps each player played.}
\label{fig:data:num-games}
\end{figure}

The complete data set consists of 476 games from 45 participants. The range of number of games played varied from 4 to 22, and has been visualized in \figref{fig:data:num-games}.

As player skill was the main focus of this research, some effort went towards ensuring balance. This was validated using the number of FPSs played ($f$), which was found to be a better indicator of the two self-reported measures. Even though there was an overall imbalance of players according to this metric, the distribution of the original population was unknown, and the range of different skills was considered acceptable.

\begin{figure}[!t]
\centering
\includegraphics[width=\figwidth]{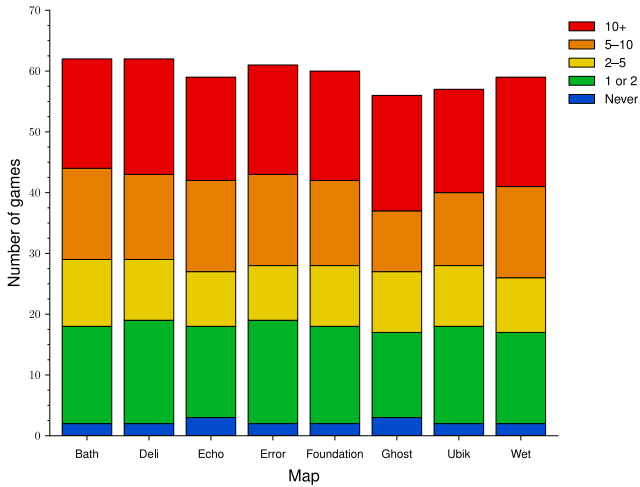}
\caption{The number of times each map was played, overlaid by the number of times played by each group in $f$.}
\label{fig:data:maps}
\end{figure}

\begin{figure}[!t]
\centering
\includegraphics[width=\figwidth]{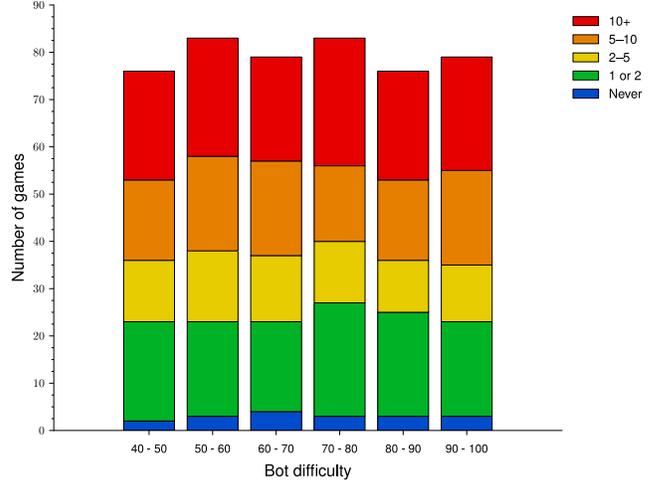}
\caption{As in \figref{fig:data:maps}, the number of games played on each difficulty, with additional grouping over $f$.}
\label{fig:data:bots}
\end{figure}

The map and bot difficulties were selected independently and uniformly at random, adjusted by the experimenter to ensure players did not have a biased experience of the game. The distribution of maps over players is also represented in \figref{fig:data:num-games}, while the maps and bot difficulties played for each skill group is shown in \figref{fig:data:maps} and \figref{fig:data:bots} respectively.

From a preliminary analysis of the first period of data, average player skill leveled out near the 6th game (more detail is provided in \secref{sec:measures}). For this study, we therefore discarded players with fewer than 8 games and ignored games played after the 16th, in order to minimize bias. This selection of data (430 games from 37 players) is highlighted in \figref{fig:data:num-games} and has been used throughout the rest of this paper.

\section{Methods}
\label{sec:method}

This section reviews the existing measures and algorithms that are used in our experiments. For our analysis of the skill metrics in \secref{sec:measures}, we present the details of the TrueSkill algorithm, and discuss some methods for evaluating similar skill metrics. Next, we introduce some techniques used to extract features from the players' input which are used in \secref{sec:features}. Finally, we present the random forest algorithm used to predict skill in \secref{sec:prediction}.

\subsection{The TrueSkill Algorithm}

TrueSkill is a widely used measure of skill in commercial games, used primarily for matchmaking, and hence serves as an important benchmark for other methods presented in this paper. The algorithm assigns unitless values, $\mu$ and $\sigma$, to players, which represent the algorithm's belief in the player's skill. The first value, $\mu$, is the current estimate, and $\sigma$ is the confidence in that estimate. Together, the two values represent a normal distribution of skill.

When two players compete, the two normal distributions can be combined to indicate the probability of a draw (the \emph{prior}). After the game, the result (the \emph{likelihood}), can be used to update the model's belief in both players. If a player, Alice ($\mu_a$, $\sigma_a$), beats Bob ($\mu_b$, $\sigma_b$), $\mu_a$ would increase, $\mu_b$ would decrease, and both values of $\sigma$ would decrease according to the following formulas:

\begin{IEEEeqnarray*}{rCl}
  \mu_{winner} & \leftarrow & \mu_{winner} + \frac{\sigma_{winner}^2}{c} \cdot V, \\
  \mu_{loser} & \leftarrow & \mu_{loser} - \frac{\sigma_{loser}^2}{c} \cdot V, \\
  \sigma_{winner}^2 & \leftarrow & \sigma_{winner}^2 \cdot (1 - \frac{\sigma_{winner}^2}{c^2} \cdot W), \\
  \sigma_{loser}^2 & \leftarrow & \sigma_{loser}^2 \cdot (1 - \frac{\sigma_{loser}^2}{c^2} \cdot W), \\
\end{IEEEeqnarray*}
where
\begin{IEEEeqnarray*}{rCl}
  c^2 & = & 2 \beta^2 + \sigma^2_{winner} + \sigma^2_{loser}, \\
  V   & = & \textsl{v}(\frac{\mu_{winner} - \mu_{loser}}{c}, \frac{\varepsilon}{c}), \\
  W   & = & \textsl{w}(\frac{\mu_{winner} - \mu_{loser}}{c}, \frac{\varepsilon}{c}). \\
\end{IEEEeqnarray*}

The functions, $\textsl{v}$ and $\textsl{w}$, dictate the update for $\mu$ and $\sigma$ respectively. This only leaves $\varepsilon$, the probability of a draw, and $\beta^2$, which is a player's performance variance. The more the performance of the players varies, the slower the values will update. A more thorough description of the workings of TrueSkill can be found in \cite{trueskill:math}.

The two values $\mu$ and $\sigma$ are usually combined to produce an ordinal value which can be used to rank players. A conservative estimate is usually used, and is given as $\mu - 3 * \sigma$ in this research.

\subsection{Evaluating Skill Metrics}

In classification problems, it is common to evaluate the model using its testing accuracy (or error rate). There are also other measures and techniques for helping to understand the model's performance. Within regression (predicting a continuous measure), the proportion of explained variance ($R^2$) is a common evaluation criteria. This measure and others, including relative absolute error (RAE) \cite{predicting-tombraider}, punish offset results and those suffering from scaling effects. The values we are comparing, however, are skill measures; measures which are ultimately used for ranking players. We therefore use Spearman's rank correlation coefficient (Spearman's $\rho$) to evaluate our models. This has the added advantage that the ranking of two different skill measures can be compared. Spearman's $\rho$ is defined as the Pearson correlation coefficient \cite{pearson:definition} between two ranked variables.

In some instances we have multiple groups of players and need to determine whether the groups are significantly different. For this situation, where the skill metrics are non-parametric, unlike a t-test, and the measures are independent, in contrast with the Wilcoxon signed-rank test, the Mann--Whitney \emph{U} test is suitable \cite{mann-whitney:definition}. In particular, given two groups of players, we can use this test to determine whether one group is statistically more skilled than the other, given different significance levels, $\alpha$.

\subsection{Complexity of Hardware Input}

A reasonable hypothesis is that skilled players use controls in a more complex way than novices. We therefore use a number of techniques to measure this complexity---some for compression of a sequence and others for analysis on a time-series. These techniques are used to extract features which are then used in Sections \ref{sec:features} and \ref{sec:prediction}.

The first two, Lempel-Ziv-Welch (LZW) \cite{lzw} and Huffman coding, can all be used for compression of data. Simple, or more predictable data, should be easier to compress, allowing these to be used to measure complexity. The first, LZW, has the advantage of being simple to implement. The algorithm is as follows:

\begin{enumerate}
  \item Initialize a dictionary with single-character strings.
  \item Find the next longest string, $W$, in the dictionary.
  \item Replace $W$ with the dictionary index.
  \item Add ($W$ + next character) to the dictionary.
  \item Go to Step 2.
\end{enumerate}

The second algorithm, designed by Huffman \cite{huffman}, constructs a Huffman tree based on probability distributions. Common characters are given smaller codes and placed towards the left of the tree. Encoding involves replacing characters with codes from the tree. If the population distribution of the characters is known, Huffman encoding is close to the theoretical minimum.

In addition to the compression techniques above, two measures of entropy are used: Shannon entropy and sample entropy. The first measures the amount of information in a given sequence:

\begin{IEEEeqnarray*}{rCl}
  H(X) & = & -\sum_{i} P(x_i)\log P(x_i).
\end{IEEEeqnarray*}

The second measure, sample entropy, based on approximate entropy \cite{sampen}, is performed on continuous data and was originally designed for physiological time-series. Independent of data length, it is potentially useful in understanding the complexity of either mouse or keyboard input.

The final complexity measure used was a discrete Fourier transform \cite{fft}. This method reveals regularities in the data and relative strengths of periodic components. Assuming complexities vary with skill, it would be interesting to see how the frequencies of the mouse input compare between users.

\subsection{Prediction Using Random Forests}

There are several techniques that could be used for predicting player skill. Previous research \cite{skill:rts:input,modeling-tombraider} successfully used SMO (Sequential Minimal Optimization), an algorithm for support vector machines \cite{smo}. However, random forests \cite{forests} were chosen for their ability to generalize well, even with a large number of features with unknown properties. A random forest also has the added advantage of being a `gray box', in that it can be used with little knowledge of its internal mechanics, but can tell us which features were the most import during training. Finally, a random forest model can be trained for each classification or regression, which can accommodate the different shapes and sizes of skill metrics.

Random forests are an ensemble method that train several trees on different subsets of the data. The MATLAB implementation used was an interface to the R implementation by Andy Liaw \textit{et al.} \cite{randomForest}. Two settings are used during training this model. The first, \texttt{ntree}, dictates how many trees to use. This was left on its default setting of 500 for all the given experiments. The second setting, \texttt{mtry}, determines how many features are sampled from when a tree is split. This variable was also left on its default setting, $\lfloor \sqrt D \rfloor$, where $D$ is the total number of features.

\section{Analysis of Skill Metrics}
\label{sec:measures}

Any research in player skill requires an understanding of the metrics used, yet there is no gold standard for measuring skill. In order to better understand how we evaluate skill, this section presents an analysis of a number of skill metrics based on our data set. For reference, these skill metrics and their notations are summarized in \tabref{tab:notations}. Although not a complete analysis, this section demonstrates how skill metrics should be understood before any analysis or prediction of skill.

\begin{table}[!t]
\renewcommand{\arraystretch}{1.3}
\caption{A summary of the skill metrics introduced in \secref{sec:measures} and their notation.}
\label{tab:notations}
\centering
\begin{tabular}{r | c | p{2.0in}}
\hline
\bfseries Name & & \bfseries Description \\
\hline \hline
Player rank & $\bar{r}$ & Mean rank ($r$) over all of an individual's games. \\
Player score & $\bar{s}$ & Mean score ($s$) over all of an individual's games. \\
TrueSkill estimate & $T$ & A TrueSkill value produced using an approximation of the TrueSkill algorithm. \\
FPSs played & $f$ & The number of FPSs the player reported they had played. \\
Hours played & $h$ & The number of hours the player reported they played per week. \\
Player KDR & $\bar{k}$ & Mean kill-to-death ratio ($k$) over all of an individual's games. \\
\hline
\end{tabular}
\end{table}

\subsection{Rank}

The winner of any game is given by a single performance metric. For chess, this is a simple win-loss-draw state. Many games use rank ($r$), where $r=1$ is the winner, $r=2$ indicates second place, and so on. $r$ is used in the TrueSkill algorithm, and is a descriptive win-loss value for games with multiple players or teams.

Rank is the value that defines performance for a single game. That makes it a logical metric to use. Although large differences in skill are ignored by $r$, it may be less easily affected by content (a win on one map will have the same value as a win on a different map). However, $r$ is still defined by the number of players on a map, and the continuum of values is limited by it. This makes it more difficult to distinguish between two players with high performance.

\begin{figure}[!t]
\centering
\includegraphics[width=\figwidth]{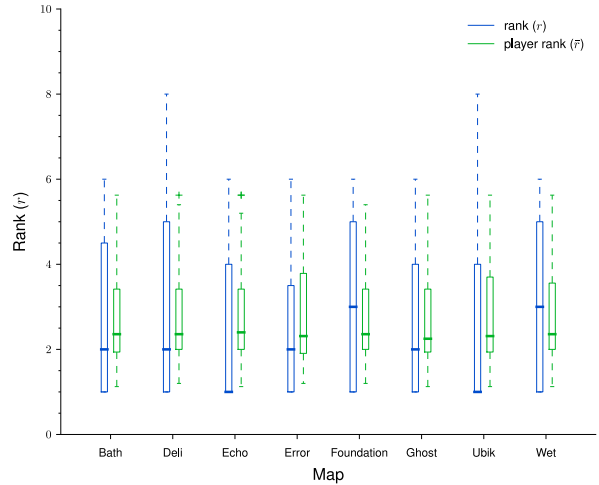}
\caption{This Tukey box plot \cite{boxplot} presents the performance metric, rank ($r$) and skill metric ($\bar{r}$) for every game, grouped by the game's map. For $r$, a lower value indicates higher performance. On average, players performed worse on the maps \emph{Foundation} and \emph{Wet}.}
\label{fig:rank:map-dependence}
\end{figure}

\begin{figure}[!t]
\centering
\includegraphics[width=\figwidth]{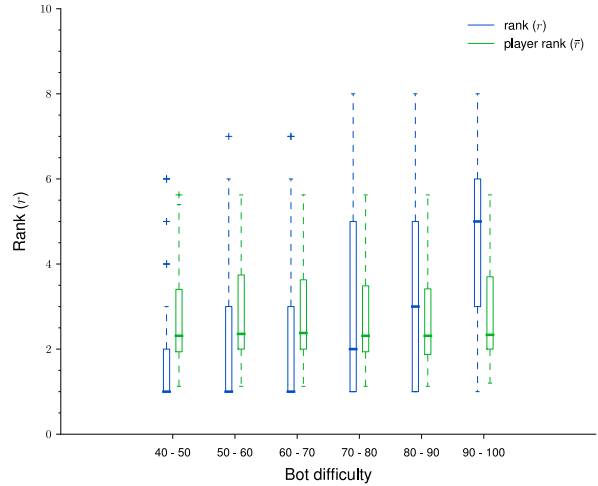}
\caption{Rank ($r$) and player rank ($\bar{r}$) presented in the same notation as \figref{fig:rank:map-dependence}, grouped instead by difficulty. Harder difficulties (e.g. 90--100) led to much lower performance.}
\label{fig:rank:bot-dependence}
\end{figure}

In our research, skill is measured over the whole task and should therefore be independent of content and difficulty. However, $r$, as a performance metric, is dependent on both map, seen in \figref{fig:rank:map-dependence}, and difficulty, \figref{fig:rank:bot-dependence}. There are two methods for averaging rank used in this paper. The first uses Bayesian updating (TrueSkill), and is discussed later. The second is obtained by taking the mean $r$ over a player's games, producing a continuous metric, player rank ($\bar{r}$).

\subsection{Score}

In order to work out the ranking of players, games often use an alternative performance measure. Racing games, for example, commonly use time. The primary goal of a deathmatch (the task in this experiment) is to accrue points. Points are accumulated by killing other players, with extra points awarded for `skillful' behavior such as assisting other players. At the end of the game, each player's rank is worked out from the number of points they have: their score ($s$). Similar scoring systems are used in other first-person shooters.

It is important to note that score can only be used as a performance metric because rank is based on it. For a different game mode or genre, a different metric should be used. \emph{Team Fortress 2}, for example, keeps a score for each player; however, as these values do not directly influence the result of the game, it is meaningless as a performance metric.

\begin{figure}[!t]
\centering
\includegraphics[width=\figwidth]{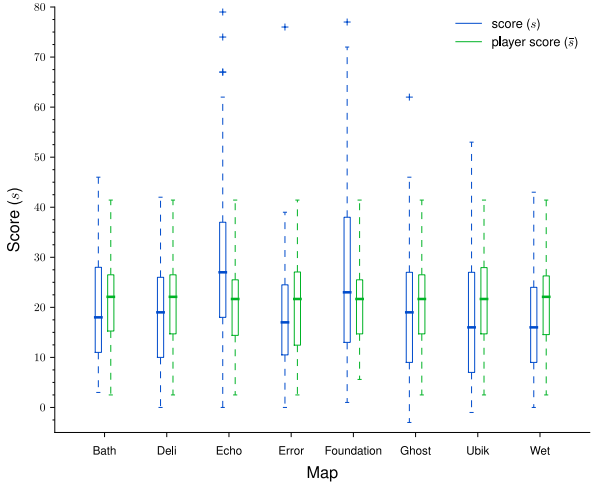}
\caption{As in \figref{fig:rank:map-dependence}, but using performance and skill metrics score ($s$) and player score ($\bar{s}$) respectively. Higher $s$ indicates higher performance.}
\label{fig:score:map-dependence}
\end{figure}

\begin{figure}[!t]
\centering
\includegraphics[width=\figwidth]{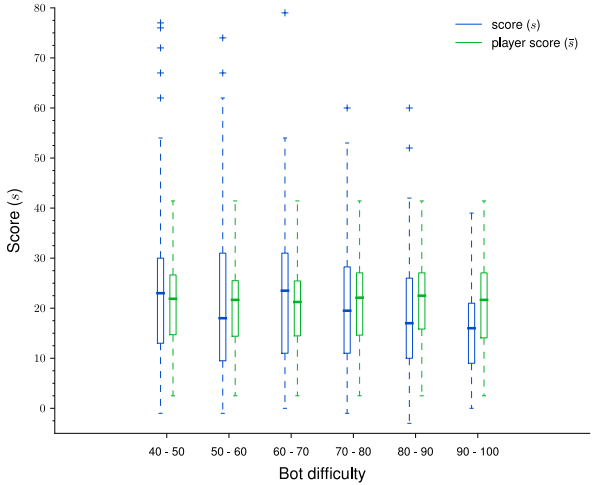}
\caption{As in \figref{fig:rank:bot-dependence}, but with score ($s$) and player score ($\bar{s}$). As with $r$, $s$ was lower for higher difficulties.}
\label{fig:score:bot-dependence}
\end{figure}

The main advantage of $s$ over $r$ is that $s$ has a much larger range of values, and is therefore more descriptive. A larger value of $s$, for instance, may imply an easier victory. On the other hand, $s$, like $r$, is dependent on content and difficulty, as seen in \figref{fig:score:map-dependence} and \figref{fig:score:bot-dependence}. The maps \emph{Foundation} and \emph{Ubik} are worth noting when comparing the two measures. For \emph{Foundation}, players tend to perform well using the performance metric $s$, but, on average, rank low. This demonstrates instances where $s$ is inflated by content. Conversely, \emph{Ubik} was a particularly hard map for players according to $s$. However, bots found it more difficult, resulting in higher ranks in \figref{fig:rank:map-dependence}.

\begin{figure}[!t]
\centering
\includegraphics[width=\figwidth]{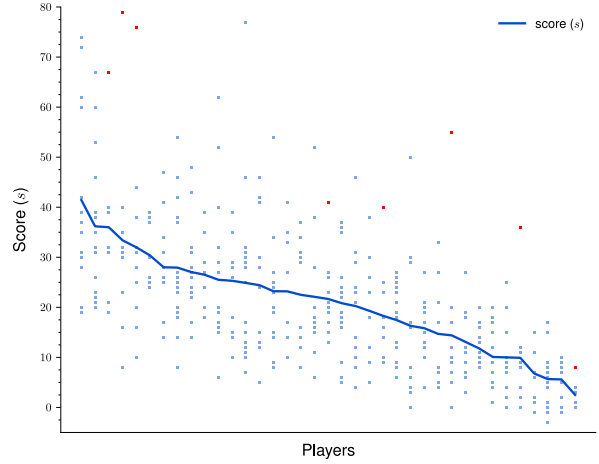}
\caption{Score ($s$) for each player, ordered by the players' mean scores ($\bar{s}$). Highlighted games indicate unexpectedly high values of $s$.}
\label{fig:score:all-players}
\end{figure}

\begin{figure}[!t]
\centering
\includegraphics[width=\figwidth]{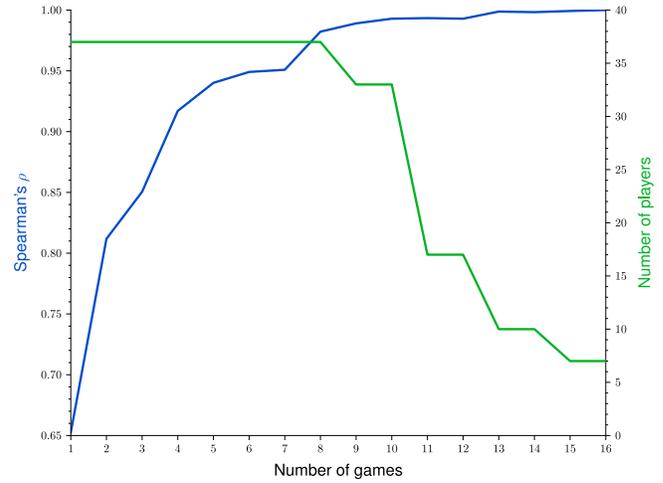}
\caption{The correlation, $\rho$ between $\bar{s}_i$ and final player score ($\bar{s}$), where $\bar{s}_i$ is calculated by averaging score ($s$) over the first $i$ games for each player.}
\label{fig:score:averaging}
\end{figure}

As with $r$, a skill metric, player score ($\bar{s}$) was produced using the mean of $s$ over all games played by a player. The $s$ and $\bar{s}$ values for each player is presented in \figref{fig:score:all-players}, illustrating the outlying values of $s$ for individuals that are accommodated for in $\bar{s}$. We were confident that some players had played enough games to obtain a reasonable skill metric, but the specific number of games required was unknown. \figref{fig:score:averaging} shows that after playing between 5 and 7 games, $\bar{s}$ starts to stabilize. The large increase in Spearman's $\rho$ between games 7 and 8 is because some players had only played 8 games.

Using $\bar{s}$, the players were separated into four bins defined in \tabref{tab:score-groups}. The limits of these groups were chosen so that there was a roughly equal number of participants in each group. These groups have been used throughout this research as a substitute for $\bar{s}$ where groups of skill are more appropriate. A directional Mann--Whitney \emph{U} test confirms that the groups are statistically different with a significance level of $\alpha = 0.005$.

\begin{table}[!t]
\renewcommand{\arraystretch}{1.3}
\caption{The different groups separated by player score ($\bar{s}$).}
\label{tab:score-groups}
\centering
\begin{tabular}{r|c|c}
\hline
\bfseries $\bar{s}$  & \bfseries Name     & \bfseries Number of Players \\
\hline \hline
$<$ 14    & Novice        & 9 \\
14--22  & Intermediate  & 10 \\
22--27  & Skilled       & 9 \\
$\geq$ 27   & Expert        & 9 \\
\hline
\end{tabular}
\end{table}

\subsection{TrueSkill Estimate}

TrueSkill is designed for multiplayer leagues, where a TrueSkill model for one player interacts with other TrueSkill models for other opponents. Unfortunately, participants in our experiment never played against each other, but against bots. In order to account for this, a slight adaptation was made to the TrueSkill algorithm.

For each game, the opponents (bots) were selected randomly from a predefined range, $b$. As we did not know the precise difficulty of each bot, we assigned a $\mu_b$ and $\sigma_b$ value to the whole range, $b$; in other words, every bot in range $b$ had the same $\mu_b$ and $\sigma_b$ values. To calculate final $\mu_b$ and $\sigma_b$ values, the TrueSkill algorithm was run over randomly selected games, updating $\mu_b$ and $\sigma_b$ with the average posteriors, $\mu$ and $\sigma$ from all bots. With these final $\mu_b$ and $\sigma_b$ values, the TrueSkill algorithm was run as normal to calculate player TrueSkill values $\mu_p$ and $\sigma_p$.

\begin{figure}[!t]
\centering
\includegraphics[width=\figwidth]{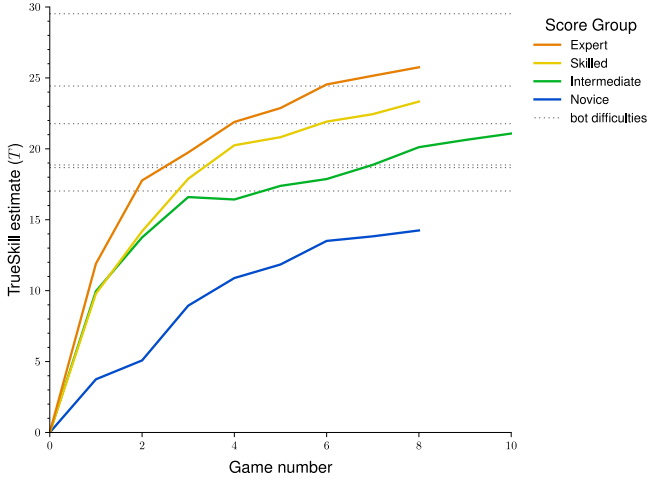}
\caption{Convergence of each player's TrueSkill estimate ($T$) over time. Bot difficulties represented by dotted lines.}
\label{fig:trueskill:convergence}
\end{figure}

Typically, a conservative estimate of skill is used for ranking: $\mu - k * \sigma$. In this research, $k$ is set to 3, i.e. $T = \mu - 3 * \sigma$. The average $T$ (TrueSkill estimate) value for each score group over time can be seen in \figref{fig:trueskill:convergence}. The dotted lines indicate the $T$ values for different bot ranges.

\begin{figure}[!t]
\centering
\includegraphics[width=\figwidth]{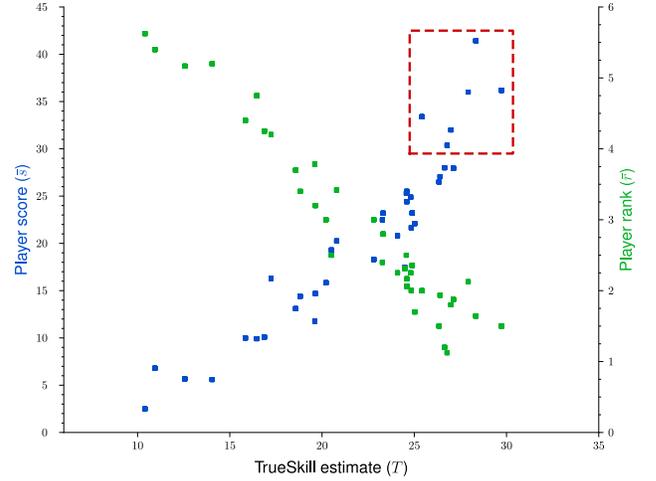}
\caption{Relationship between the skill metrics TrueSkill estimate ($T$), player score ($\bar{s}$) and player rank ($\bar{r}$). Highlighted players have much higher relative values of $\bar{s}$ than their equivalent $T$ or $\bar{r}$.}
\label{fig:trueskill:relationship}
\end{figure}

To our knowledge, the TrueSkill algorithm has not been applied to single-player content before, or to simulated multiplayer, where players compete against bots. Although \figref{fig:trueskill:relationship} shows that $T$ generally agrees with both metrics, $\bar{r}$ and $\bar{s}$, we do not know how valid this method is. It may also be that $T$ values for players, some of whom played as few as 8 games, did not fully converge. In addition, $\bar{s}$ discriminates between the higher-end players ($T > 25$) more effectively than $T$ or $\bar{r}$.

\subsection{Self-Reported Measures}

Asking players about their gaming experience is common in related research \cite{corpus:mario}. It can serve to put research into context, and is very easy to collect. In commercial games, players are commonly asked to select a difficulty setting. However, players are poor estimators of their own skill \cite{self-assessment}. This research therefore explores two objective criteria for reporting player experience, hours played ($h$) and FPSs played ($f$).

\begin{figure}[!t]
\centering
\includegraphics[width=\figwidth]{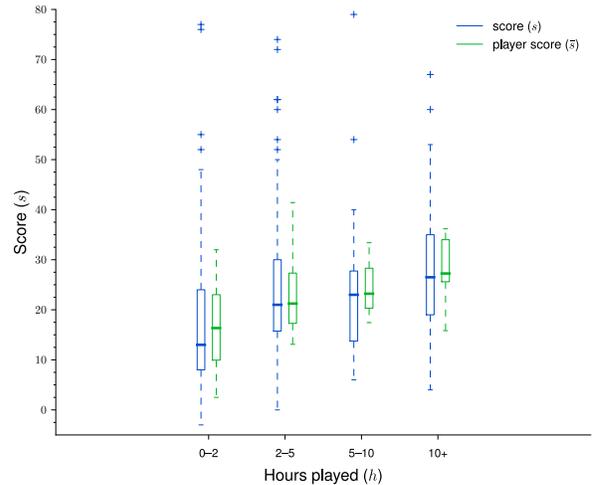}
\caption{The distribution of score ($s$) and player score ($\bar{s}$) for each hours played ($h$) group.}
\label{fig:hours:skill-measures}
\end{figure}

The number of hours that someone plays games for may be indicative of playing behavior. It may not, however, relate well to skill. \figref{fig:hours:skill-measures} illustrates how this value compares with a performance measure, $s$, and a skill measure, $\bar{s}$. In addition to the low correlation between the groups (\tabref{tab:skill-metrics}), there is significant overlap of skill between the groups, and some players from~$h =$~2--5 have a higher score than those in higher skill categories. Indeed, using a directional Mann--Whitney \emph{U} test with a significance level of $\alpha = 0.025$, there was not sufficient evidence to state that any group was statistically greater than its previous group. There were, however, not enough players for the pair of groups~$h =$~5--10 and~$h =$~10+ to make any conclusions.

\begin{figure}[!t]
\centering
\includegraphics[width=\figwidth]{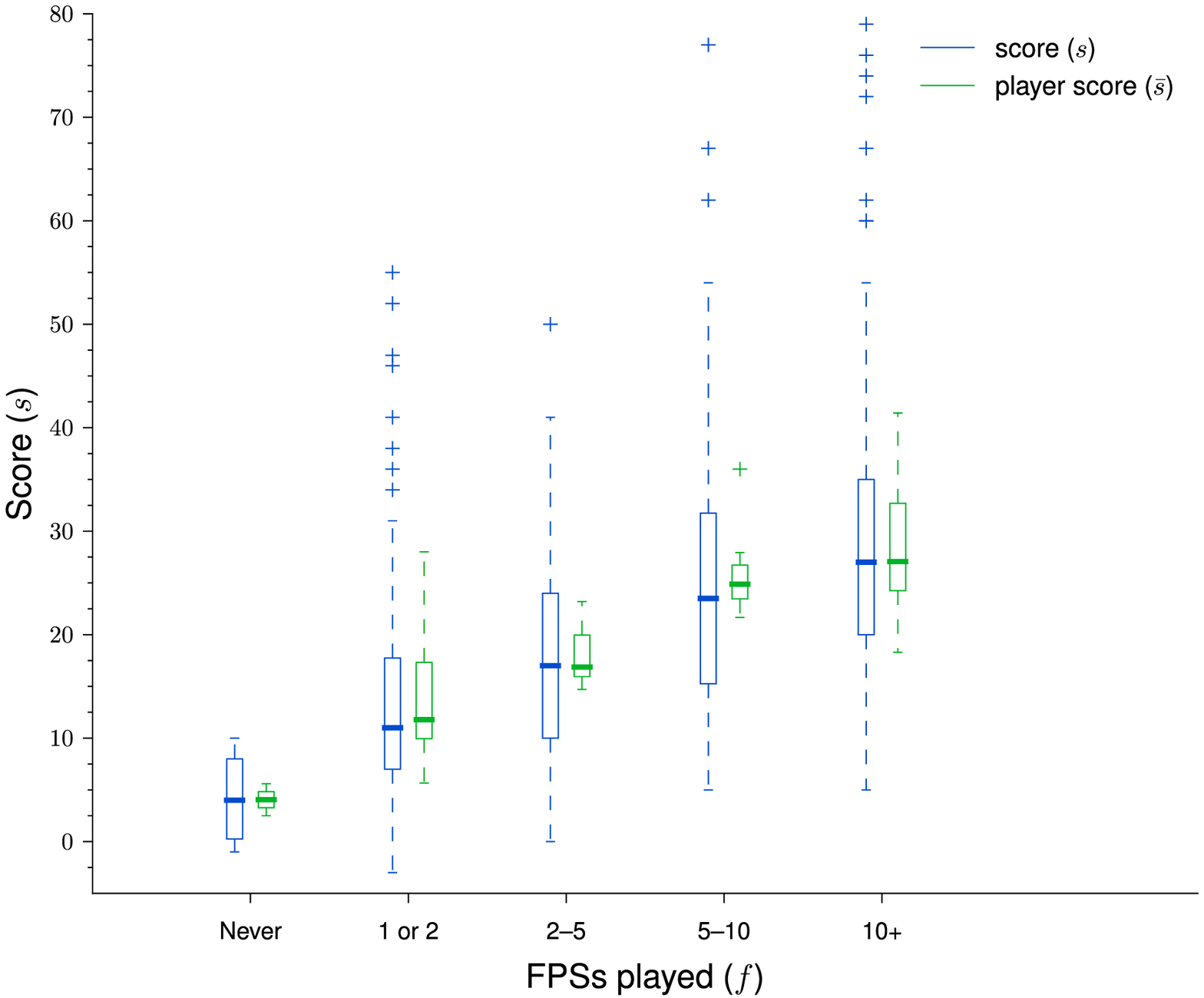}
\caption{The distribution of score ($s$) and player score ($\bar{s}$) for each FPSs played ($f$) group.}
\label{fig:fps:skill-measures}
\end{figure}

The second metric, $f$, consists of 5 categories and attempts to take into account the user's entire gaming history and exclude time spent playing other genres of game, such as role-playing games. Again, a comparison between $f$ and $\bar s$ is shown in \figref{fig:fps:skill-measures}. Although more closely correlated to skill, there were a few players in the second category,~$f =$~1~or~2, who had more experience than could be described by this measure. As done with $h$, a directional Mann--Whitney \emph{U} test was performed between each adjacent measure. There were not enough players for the pair~$f =$~Never and~$f =$~1~or~2. Between the other pairs, only~$f =$~5--10 was found to be greater than its predecessor,~$f =$~2--5 with a significance level of~$\alpha = 0.005$.

\begin{table*}[!t]
\renewcommand{\arraystretch}{1.3}
\caption{Correlation (Spearman's $\rho$) of all skill metrics, where values $>$ 0.9 and $<$ -0.9 are highlighted.}
\label{tab:skill-metrics}
\centering
\begin{tabular}{r||*{10}{c}}
\hline
& $\bar{s}$ & $\bar{r}$ & $\bar{k}$ & $\bar{a}$ & $T$ & $\bar{d}$ & $f$ & $h$ \\
\hline \hline
$\bar{s}$& - &\bfseries -0.9103 & 0.8770 & 0.6752 & \bfseries 0.9614 &-0.1156 & 0.7699 & 0.5005 \\
$\hat{r}$& - & - &-0.8476 &-0.6811 &\bfseries -0.9545 & 0.3069 &-0.6768 &-0.4671 \\
$\bar{k}$& - & - & - & 0.5071 & 0.8537 &-0.4240 & 0.6584 & 0.4557 \\
$\bar{a}$& - & - & - & - & 0.6432 &-0.1390 & 0.4761 & 0.4126 \\
$T$      & - & - & - & - & - &-0.1620 & 0.7219 & 0.4999 \\
$\bar{d}$& - & - & - & - & - & - &-0.0434 &-0.1724 \\
$f$      & - & - & - & - & - &- & - & 0.3533 \\
$h$      & - & - & - & - & - &- & - & - \\
\hline
\end{tabular}
\end{table*}

\subsection{Community Measures}

The gaming community will often use game statistics to evaluate and compare players. These are designed to give a better understanding of each player's strengths and weaknesses, but are often specific to the game genre they are used for, such as \emph{actions-per-minute} in \emph{StarCraft}.

Kill-to-death ratio ($k$), often abbreviated KDR, and accuracy ($a$) are two performance measures that are specific to first-person shooters. The first, $k$, represents the number of kills the player made against the number of times they were killed themselves. The second, $a$, is the hit ratio of the player; the number of times they hit opponents versus the number of shots they fired. Player averages have been calculated for both of these values, $\bar{k}$ and $\bar{a}$ respectively. A third measure, average number of deaths for a player ($\bar{d}$), has been included in \tabref{tab:skill-metrics} for comparison.

\begin{figure}[!t]
\centering
\includegraphics[width=\figwidth]{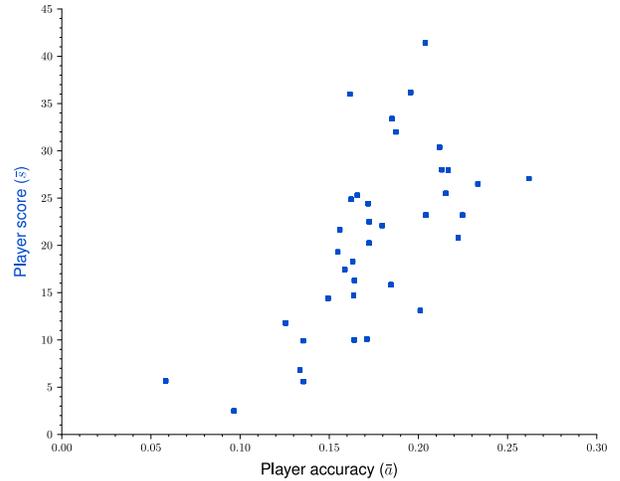}
\caption{The relationship between player accuracy ($\bar{a}$) and player score $\bar{s}$.}
\label{fig:accuracy:skill-measures}
\end{figure}

The relationship between $\bar{a}$ and $\bar{s}$ has been visualized in \figref{fig:accuracy:skill-measures}. It can be seen from this graph that although greater skill may imply greater accuracy, there is less difference of accuracy between the more skilled players. This may imply that accuracy is an ability more quickly mastered. However, the correlation between $\bar{a}$ and $\bar{s}$ is too low to make concrete conclusions about their relationship for so few players.

In summary, the three skill metrics $\bar{s}$, $\bar{r}$ and $T$ rank players very similarly. The two self-reported measures, $f$ and $h$, on the other hand, were found to be insufficient for our purposes. Equally, the community-based metrics, $\bar{k}$ and $\bar{a}$, may describe skill given a different task, but, for this experiment, are more likely to describe play style. Given that $T$ is only an estimate of TrueSkill, $\bar{s}$, as the more descriptive of the three metrics, is used for the rest of this paper.

\section{Player Input Feature Analysis}
\label{sec:features}

Using the methods presented in \secref{sec:method} and previous work \cite{skill:hci:lowlevel}, 174 global features were extracted from the keyboard and mouse events of each game\footnote{The complete list of features and their associated groups can be found on the website.}. These features are grouped and analyzed in this section in order to better understand player input and how it relates to skill.

Three different schemes, summarized in \tabref{tab:groups}, were used to group the features. By grouping these features, we can start to see how different types of player input are affected by skill. While the groups of each scheme were designed to be mutually exclusive, some features could not be categorized, so are left ungrouped, and were not used in analysis.

\begin{table}[!t]
\renewcommand{\arraystretch}{1.3}
\caption{Feature groups used within this research}
\label{tab:groups}
\centering
\begin{tabular}{r||l|c}
\hline
\bfseries Group name  & \bfseries Description     & \bfseries Features \\
\hline \hline
Keyboard              & From keyboard events      & 83 \\
Mouse                 & From mouse movement events& 66 \\
Clicks                & From mouse clicks         & 14 \\
Ungrouped             & -                         & 11 \\
\hline \hline
Event Frequency & Frequency of events over the game & 31 \\
Complexity & Complexity of input & 75 \\
Kinetics & Describing how the player or mouse moves & 19 \\
Ungrouped & - & 49 \\
\hline \hline
Context-Free & No prior knowledge of game required  & 78 \\
Dependent & Some knowledge of game semantics needed & 96 \\
\hline
\end{tabular}
\end{table}

\subsection{Hardware: Keyboard, Mouse Movement and Clicks}

The first set of groups separates features according to which input device generated the events. As one of the first obstacles to playing a game, use of the input devices is likely to contribute to skill. In addition, different types of games may have different dependencies on each of the devices.

The features extracted from the \emph{Keyboard} events concerned the complexity of the input or the frequency with which they were pressed. Some of these features were based specifically on the movement keys, which allow the player to move around. A number of mouse movement events have already been used in related HCI research \cite{skill:hci:lowlevel}, and these formed the basis for the \emph{Mouse} features. Mouse \emph{Clicks}, having been used less in the literature and far more simple in nature, had the fewest features. One set of features was created using knowledge of both mouse and keyboard and, as such, did not fall into one single category. These were ignored for this particular grouping.

\begin{figure}[!t]
\centering
\includegraphics[width=\figwidth]{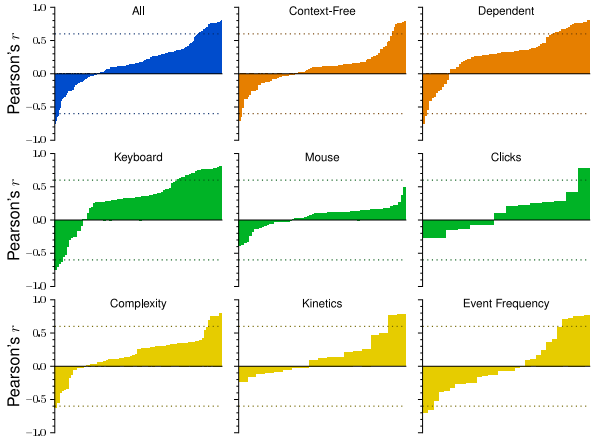}
\caption{Pearson correlation coefficient for each feature to player score ($\bar{s}$), grouped by feature group and ordered by correlation. Dotted lines indicate correlation of $\pm 0.6$.}
\label{fig:groups:correlation}
\end{figure}

\begin{figure}[!t]
\centering
\includegraphics[width=\figwidth]{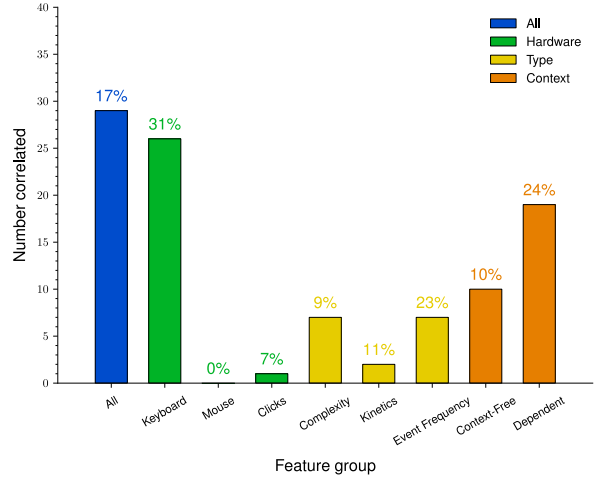}
\caption{Number of features strongly correlated to player score ($\bar{s}$) for each feature group.}
\label{fig:groups:summary}
\end{figure}

The Pearson correlation coefficient was calculated for each feature with respect to $\bar{s}$, chosen as a major index of skill, and presented in \figref{fig:groups:correlation}, grouped by feature group. The number of these with a strong correlation (defined here as 0.6, slightly greater than that suggested in previous work \cite{pearson:limit}) has been summarized in \figref{fig:groups:summary}. Although \emph{Keyboard} contains the most features, it was also one of the more interesting groups, as most features were correlated in some way. The \emph{Mouse} group, on the other hand, correlated significantly less with skill overall. This contrasts previous work in HCI, in which mouse features played a key role \cite{skill:hci:lowlevel}. \emph{Clicks} were also generally uncorrelated to skill, the most interesting being the LZW complexity of a player's clicks, with a correlation of only 0.418.

\subsection{Type: Event Frequency, Complexity and Kinetics}

The second grouping scheme is slightly less obvious, in that features are grouped according to what type of input they describe. Some features, for instance, describe the kinetic motion of the mouse, whereas others describe how complex a user's input was (according to the algorithms presented in \secref{sec:method}). These groups allow us to see what types of player input are most relevant to skill. Unfortunately, there were 49 ungrouped features which did not fall into any of the three groups within this category.

There were a number of \emph{Complexity}-based features that correlated to skill. In particular, these described how complex a player's keyboard input was. For example, the LZW complexity of the four movement keys (forward, left, right and back) correlates highly with skill (Pearson's $r = 0.799$). Skilled players had a higher LZW value, implying their skill is more complex according to the LZW algorithm.

The \emph{Kinetics} group was much smaller than its counterparts. The most interesting features, corresponding to $r \approx 0.48$, include the number of times the player changed the x-direction of the mouse and the average angle of change in a player's movement.

\emph{Event Frequency} described how often a player generated events with the input devices. Several features of this group correlated well with skill, as illustrated in \figref{fig:groups:summary}. In general, the higher a player's skill, the greater the number of presses, and the longer each key was pressed.

\subsection{Context: Free and Dependent}

In an ideal scenario, data collection could be done independently of each game. By splitting the features into those that require some prior knowledge about the game (e.g. the user pressed a key that moves the player forward), and those that do not (e.g. the user pressed the `w' key), we start to understand how independent the features are from the game. This category had the most balanced grouping out of each set. The \emph{Dependent} group comes out on top, as seen in \figref{fig:groups:summary}. This was expected, given that this group was allowed to know more about the game. On the other hand, features extracted from the keyboard without knowing anything about the game still contained some information about skill. The length of time any two keys were pressed at once, for instance, had a correlation to $\bar{s}$ of 0.780.

\begin{figure}[!t]
\centering
\includegraphics[width=\figwidth]{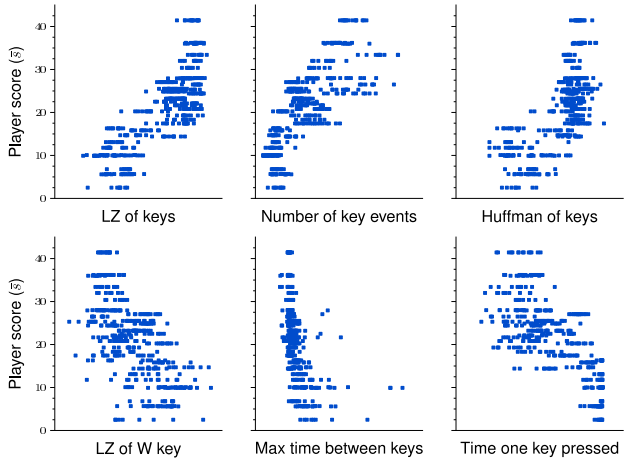}
\caption{How different features correlate to skill in different ways.}
\label{fig:correlation:types}
\end{figure}

Having found the strongest correlations for each of the groups, we identified 6 distinct types of correlation, which are presented in \figref{fig:correlation:types}.

\subsection{Player Learning}

\begin{figure}[!t]
\centering
\includegraphics[width=\figwidth]{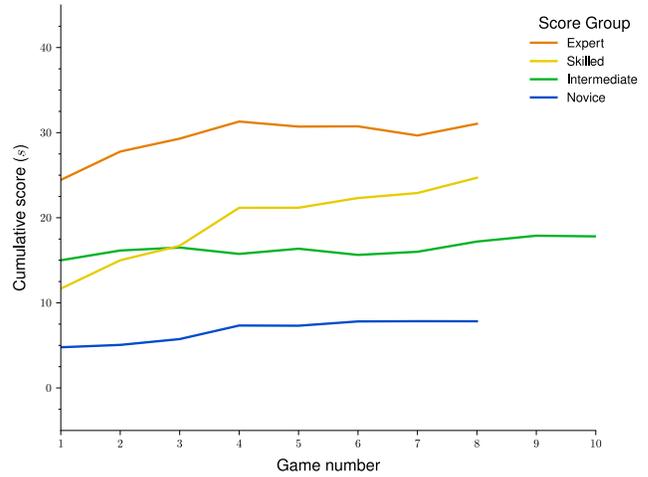}
\caption{Cumulative average score ($s$) over several games for each score group.}
\label{fig:learning:scores}
\end{figure}

The cumulative average score for each score group has been presented in \figref{fig:learning:scores}. There is a notable increase in average performance over the first few games for groups \emph{Skilled} and \emph{Expert} which is less visible in the other two groups. Given that only one person had played \emph{Red Eclipse} before, this is consistent with previous research that found skilled players learned faster \cite{skill:surgery}.

\begin{figure}[!t]
\centering
\includegraphics[width=\figwidth]{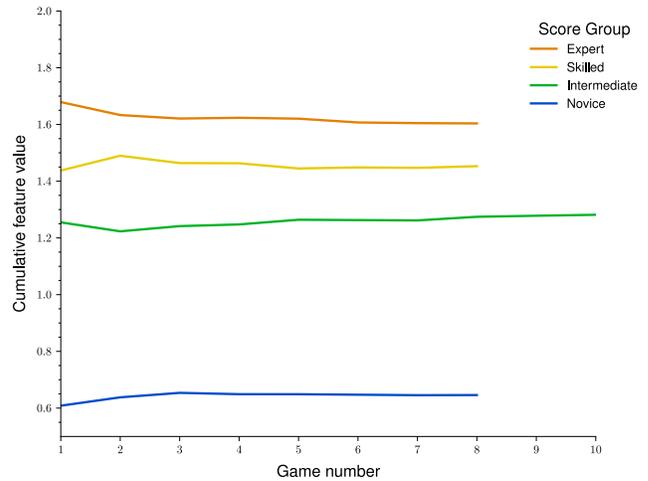}
\caption{Cumulative average value for a feature over several games for each score group.}
\label{fig:learning:feature}
\end{figure}

Selecting a feature that was particularly highly correlated with player score (the average number of keys pressed at once), we plot the cumulative average value for this over successive games in \figref{fig:learning:feature}, again grouping by score group. In contrast to \figref{fig:learning:scores}, there is much less variation in value over several games. This suggests the feature values extracted from the input are more stable than performance metrics.

\section{Skill Prediction}
\label{sec:prediction}

This section presents how a player's skill can be predicted from their input to a game. The experiments presented include predicting different classes of skill, predicting continuous skill measures and finally attempting to learn from smaller sections of gameplay. Each experiment used the random forests presented in \secref{sec:method}, and used 5-fold cross-validation.

\subsection{Predicting a Skill Category}
\label{sec:prediction:classification}

Categories of player can be used to get a general idea of how skillful players are. \emph{StarCraft}, for instance, groups players into leagues, where players in the same league are generally comparable in skill \cite{skill:rts:input}. The score groups introduced in \secref{sec:measures} are therefore used to construct a classification model.

\begin{figure}[!t]
\centering
\includegraphics[width=\figwidth]{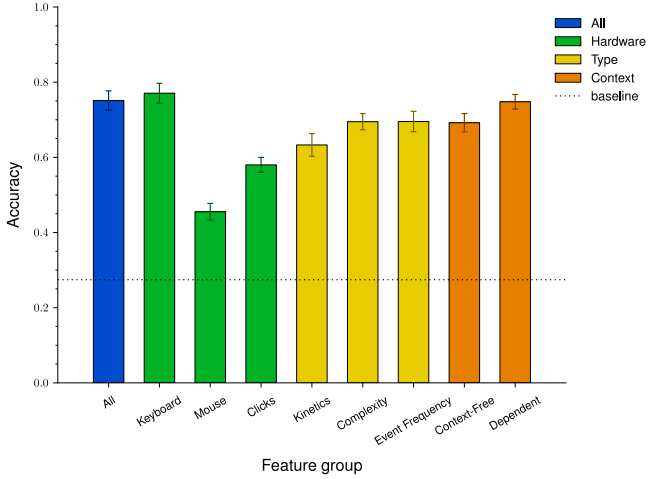}
\caption{Accuracy of random forest trained to predict a player's score group using different feature groups. Error bars indicate standard error of each model.}
\label{fig:classification:score}
\end{figure}

The average accuracy for such a model trained on the different feature groups is presented in \figref{fig:classification:score}. An average accuracy of \textbf{77.1\%} is achieved by training on \emph{Keyboard} features, significantly higher than the majority class baseline of 27.4\%. The confusion matrix of this model is given in \tabref{tab:classification:confmat}, and shows that many mistakes (77.1\% of all misclassifications) are in neighboring classes. The \emph{Intermediate} group was, however, the most difficult to predict.

\begin{table}[!t]
\renewcommand{\arraystretch}{1.3}\caption{How each game was classified for a random forest trained to predict groups of player score ($\bar{s}$).}
\label{tab:classification:confmat}
\centering
\begin{tabular}{r||*{4}{c|}c}
\hline
& \bfseries Novice & \bfseries Intermediate & \bfseries Skilled & \bfseries Expert & \\
\hline \hline
\bfseries Novice        & \bfseries 93 &  5 &  0 &  0 &  98 \\
\bfseries Intermediate  & 15 & \bfseries 64 & 19 & 20 & 118 \\
\bfseries Skilled       &  0 & 19 & \bfseries 57 & 30 & 106 \\
\bfseries Expert        &  0 &  9 & 14 & \bfseries 85 & 108 \\
\hline
              & 108&  97&  90& 135& \\
\hline
\end{tabular}
\end{table}

\begin{figure}[!t]
\centering
\includegraphics[width=\figwidth]{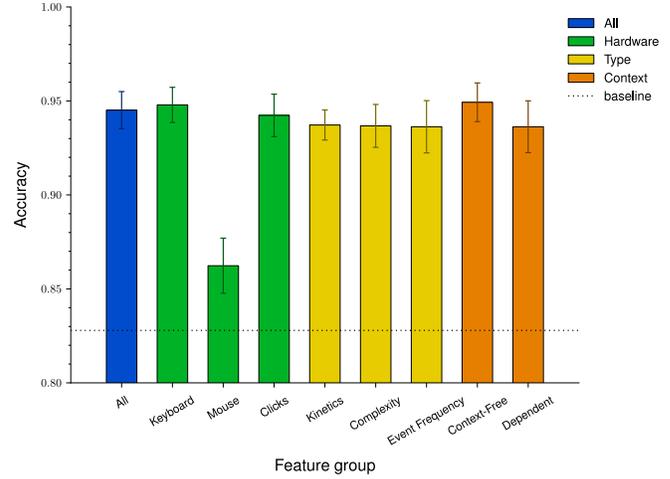}
\caption{Accuracy of random forest trained to detect \emph{Novice} players using different feature groups.}
\label{fig:classification:binary}
\end{figure}

For some applications, it is often sufficient to be able to distinguish between two kinds of players: those who have never played before, and those who have. For this binary classification, we split the data into two groups: \emph{Novice} and all others. As shown in \figref{fig:classification:binary}, the \emph{Context-Free} group achieves an accuracy of \textbf{94.9\%}, whereas the worst group, \emph{Mouse}, performed at 86.2\%.

\subsection{Predicting Skill Measures}

Most metrics of skill use a continuous measure, allowing detailed comparisons between different players. A regression model would allow these continuous values to be predicted for each player, but has not been studied in the literature as thoroughly. Predicted values are represented in this research with a hat (e.g. $\hat{\bar{s}}$ is the prediction of $\bar{s}$).

\begin{figure}[!t]
\centering
\includegraphics[width=\figwidth]{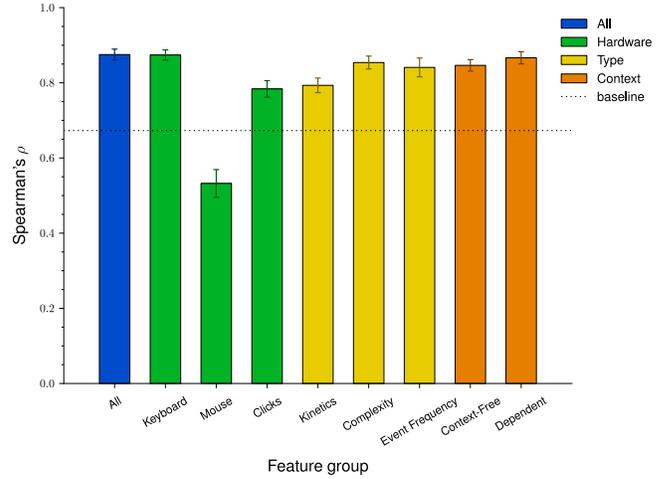}
\caption{Performance (Spearman's $\rho$) of models trained to predict player score ($\bar{s}$). Baseline indicates $\rho$ between score ($s$) and $\bar{s}$.}
\label{fig:regression:summary}
\end{figure}

We constructed several models to predict $\bar{s}$ using different feature groups, measuring the performance for each model using Spearman's $\rho$. The performances for these models are summarized in \figref{fig:regression:summary}. The comparative baseline for this experiment is to use the player's performance, $s$ (which can be collected after one game), as a substitute for the skill measure, $\bar{s}$. $\bar{s}$ is successfully predicted with \textbf{$\rho = 87.4$}, notably higher than $s$, which has a correlation of only $\rho = 67.3$.

\begin{figure}[!t]
\centering
\includegraphics[width=\figwidth]{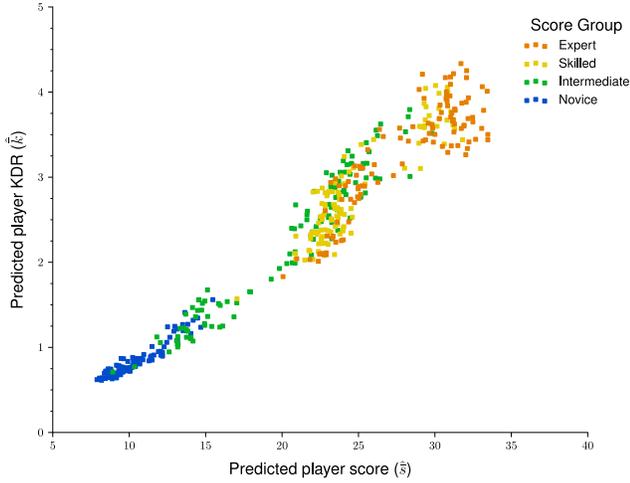}
\caption{Relationship between predicted player score ($\hat{\bar{s}}$) and predicted player KDR ($\hat{\bar{k}}$), colored by score group.}
\label{fig:regression:predicted}
\end{figure}

We visualized the average predicted values of player score ($\bar{s}$) and player KDR ($\bar{k}$) for each game ($\hat{\bar{s}}$ and $\hat{\bar{k}}$) in \figref{fig:regression:predicted}. It is clear that the two models agree with each. It also clusters the games into three groups, around $\hat{\bar{s}} < 19$ and $\hat{\bar{s}} > 27$. From this graph, it seems particularly difficult for the model to distinguish between the two highest skilled groups. It may be that the clusters created here related to both skill and player style \cite{style:lda}.

\subsection{Prediction Convergence Rate}

The features used up to this point were all extracted from the entire three minutes of gameplay. However, in order to explore how soon a player's skill could be predicted, the same features were extracted from smaller segments of the game. In addition to the full 180~s segment already used, data was extracted from the first $t$~s of the game, where several values of $t$ were selected from between 1~s and 120~s.

\begin{figure}[!t]
\centering
\includegraphics[width=\figwidth]{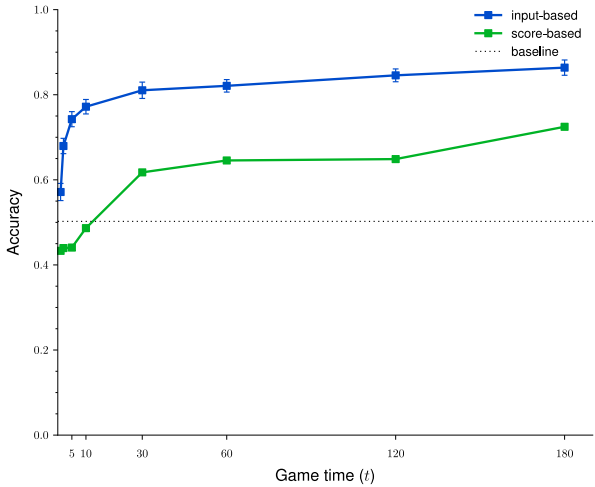}
\caption{How fast a classification model is able to predict binary score group. Dotted line indicates accuracy guessing the majority class.}
\label{fig:speed:classification}
\end{figure}

\begin{figure}[!t]
\centering
\includegraphics[width=\figwidth]{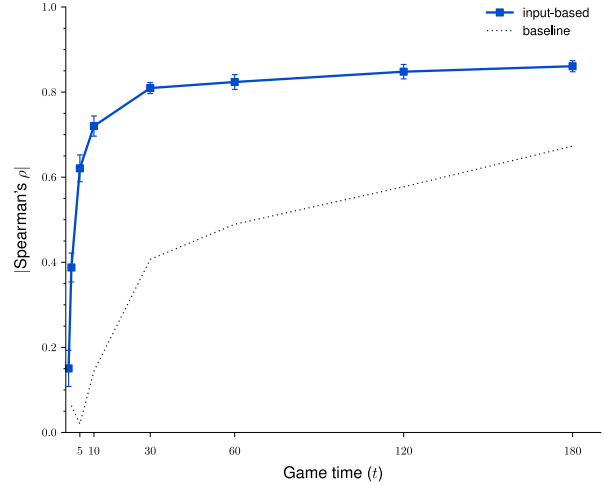}
\caption{How fast a regression model is able to predict player score ($\bar{s}$). Baseline indicates correlation of the current score at $t$.}
\label{fig:speed:regression}
\end{figure}

Splitting the players into two roughly equally-sized groups, \emph{Novice} and \emph{Intermediate} players in one group, the \emph{Skilled} and \emph{Expert} players in the other, we trained the model on the different segment sizes. The result of this is presented in \figref{fig:speed:classification} and compared to a model trained using score ($s$) as a feature. We performed the same test for a regression model, predicting $\bar{s}$ for each segment of the game. The performance of this is compared to how well the current score correlates to $\bar{s}$ in \figref{fig:speed:regression}. Not only are the input-based models more accurate than their baselines, they start to converge in a very short time (e.g. \textbf{$t = 10$~s}).

\section{Discussion}
\label{sec:discussion}

This section will discuss the implications and limitations of each contribution in turn, and finally outline future work that this research leaves open.

Our data set, presented in detail in \secref{sec:data}, is potentially useful to anyone delving into player input, particularly where two distinct input devices are required. Although our research did not offer promising results with regards to mouse input, it may be that there still exist features of mouse input that can describe a player, their style, or even their skill level. Some information about game events is also available, although limited with regards to enemies and player positions. The player experience feedback has also been left unexplored, leaving open an entirely different subject of research. The biggest limitation with this data set, one that directly affects this research, is the lack of expert players for this specific game. In the real world, expert players have more experience than those that took part in our experiment. There are also too few games per player to make adequate conclusions about learning or to explore how metrics change over time.

To our knowledge, skill metrics have not been analyzed in this way before. \secref{sec:measures} therefore provides a framework and baseline for comparing skill measures in other games. We explored two player-reported metrics ($f$ and $h$) and showed that they were inappropriate for analysis of skill. Finally, the Bayesian-averaged $T$ values, while estimates, were shown to correlate well with other skill metrics. This may indicate that similar methods can be applied to single-player games to measure skill or difficulty.

During our analysis of the features, we found that the keyboard was the most descriptive input device of skill. The mouse features, on the other hand, was very weakly correlated to player skill. While useful in previous research \cite{skill:hci:lowlevel}, it may simply be too random for use in this application\footnote{Any findings in this research are limited to the types of features extracted. There may yet be other features of mouse movement that correlate well with skill.}. We also showed that even though knowledge of the game was preferred when extracting features, there were features that correlated with skill which required no knowledge of the game. Using this, models based on a game-independent approach may be implemented externally to a game. On the other hand, the features extracted were limited to input features. As such, the predictive models may be limited to predicting skill at using the input, or mechanical dexterity, discounting other aspects of skill.

There are several key differences between the prediction done in this research and that in previous work \cite{skill:fps:input,skill:rts:input}. The first is the prior analysis of the skill metrics, which lends more credence to the results. Secondly, we showed that skill metrics such as average score ($\bar{s}$) could be predicted relatively accurately---more so than using a performance metric. And finally, this could be achieved after only a few seconds of gameplay. Using this model, matchmaking algorithms could initialize skill values for players, and then switch to another slower, but more reliable, model after a few games.

We defined our task as the average skill at deathmatch over a preselected number of maps. This meant that our `ground truth' was the average position of the player compared to other players, $\bar{r}$. If the task changed, however, to a different game mode, or to a different game, the ground truth would undoubtedly change, and as such, the meanings of each of these skill measures. In addition, although each player in our data set experienced a well-balanced proportion of content, traditional games may offer more content to the player, and a player may have a preference for particular maps, skewing a metric such as $\bar{s}$. As such, the different averaging techniques should be compared to account for differences in content.

The most obvious next step with this research is to show how these techniques can be applied. An obvious example, as already mentioned, is matchmaking. Would using a rapid model presented here help improve matchings over the first few games in a matchmaking system? And similarly, in single-player games, can a rapid model reliably select the difficulty for players, removing the need for players to learn what the developer means by `normal' or `hard'?

Many of the features we collected are relevant to all PC-based first-person shooters. Two possible extensions on this work are either generalizing to other games in the genre or attempting to predict skill on console devices. Difficulties may start to arise with the former when a game's pace changes. The \emph{Counter-Strike} series, for example, are much slower paced than \emph{Red Eclipse}, and \emph{Team Fortress 2} lets players compete as different classes, each with different styles of play. Console games, on the other hand, control player movement using analogue sticks, which may require completely different types of features.

\section{Conclusions}
\label{sec:conclusions}

This research has provided a strong foundation for skill capture in video games by presenting some methods for understanding metrics used. More specifically, we demonstrated that skill could be predicted reliably after only 10s of gameplay (see \figref{fig:speed:regression}).

The applications for this research can be directly applied to matchmaking and DDA systems, potentially improving player satisfaction in the short-term. However, the models will need to be further refined or adapted when applied to other domains or when using different input devices. The area of research that we intend to explore next is skill capture in single-player games, applying the same methods presented here, and showing how they can be used to improve DDA.



\bibliographystyle{IEEEtran}
\bibliography{IEEEabrv,refs}

\end{document}